\documentclass[amsmath,amssymb,superscriptaddress,showpacs,preprint]{revtex4-1}

\usepackage{graphicx}
\usepackage{dcolumn}
\usepackage{bm}
\usepackage{amsmath}
\usepackage{amssymb,amsthm}
\usepackage{color}
\usepackage{epstopdf} 
\usepackage{subfig}

\usepackage{epstopdf}
\usepackage{mathbbol}

\usepackage{xcolor,soul} 

\begin{document}



\title{Cell population heterogeneity driven by stochastic partition and growth optimality}

\author{Jorge Fernandez-de-Cossio-Diaz}
\email{cossio@cim.sld.cu}
\affiliation{Systems Biology Department, Center of Molecular Immunology, Havana, Cuba}
\affiliation{Group of Complex Systems and Statistical Physics, Department of Theoretical Physics, University of Havana, Physics Faculty, Cuba}
\author{Roberto Mulet}
\email{mulet@fisica.uh.cu}
\affiliation{Group of Complex Systems and Statistical Physics, Department of Theoretical Physics, University of Havana, Physics Faculty, Cuba}
\affiliation{Italian Institute for Genomic Medicine, IIGM, Torino, Italia}
\author{Alexei Vazquez}
\email{a.vazquez@beatson.gla.ac.uk}
\affiliation{Cancer Research UK Beatson Institute, Glasgow, United Kingdom}
\affiliation{Institute for Cancer Sciences, University of Glasgow, Glasgow, United Kingdom}

\date{\today}

\begin{abstract}
A fundamental question in biology is how cell populations evolve into different subtypes based on homogeneous processes at the single cell level. Here we show that population bimodality can emerge even when biological processes are homogenous at the cell level and the environment is kept constant. Our model is based on the stochastic partitioning of a cell component with an optimal copy number. We show that the existence of unimodal or bimodal distributions depends on the variance of partition errors and the growth rate tolerance around the optimal copy number. In particular, our theory provides a consistent explanation for the maintenance of aneuploid states in a population. The proposed model can also be relevant for other cell components such as mitochondria and plasmids, whose abundances affect the growth rate and are subject to stochastic partition at cell division.
\end{abstract}

\maketitle


\clearpage



\clearpage

\section*{Introduction}

It is generally believed that, in simple homogeneous environments, only one competitor class can be sustained on a single resource \cite{hardin1960s}.
The intuition is that in the long run, a fittest competitor class outgrows the rest.
However, this contradicts the existence of structured and heterogeneous communities, and in particular their emergence from initially clonal populations \cite{Altschuler10}.
One explanation to this contradiction rests in the idea that evolution in simple environments is a sequence of selective sweeps where dominant clones are regularly replaced by fitter descendants \cite{Kimura1983Neutral}.
In this case diversity is a temporary state in the transition from a dominant clone to another.
The validity of this formulation is restricted to a regime where replication errors giving rise to fit descendants are rare.
In the opposite extreme, the quasi-species model applies to very large populations with frequent mutations \cite{eigen1989molecular}, such as viruses.
In this case, a population is predicted to form a cloud around a fitness peak (the so-called ``quasi-species'' \cite{eigen1989molecular}), unless the mutability exceeds an ``error threshold'', in which case individuals drift randomly on the fitness landscape \cite{Nowak1989finiteQuasispecies}.
Since the distribution is uni-modal or uniform, in this case there are no clearly defined sub-types within the population.
Furthermore, the high mutation rates of viruses are not common in other types of cell, and this restricts the applicability of the model.
On the other hand, mathematical descriptions of chemostat experiments predict that diversity cannot be maintained unless cells engage in cross-feeding \cite{Pfeiffer2004crossfeeding}, are subject to product inhibition \cite{wortel2016scirep}, rate-yield trade-offs \cite{beardmore2011n,Pfeiffer2001}, or there are periodic variations in the dilution rate of the chemostat \cite{Butler1985mathchemostat}.

Despite these difficulties to explain the emergence of multiple cell types, experiments supporting the phenomenon are ubiquitous.
Clonal bacterial populations can diverge into multiple phenotypic clusters in the chemostat \cite{maharjan2006s}.
Tissues of multicellular organisms are composed of a hierarchy of genetically identical cells with different phenotypes that maintain a stable coexistence \cite{luksza2017neoantigen,stamp2018predominant}.
Resistance to cancer therapy is related in part to the heterogeneity of cancer cells before treatment \cite{Meacham13,Dick08,wang2016clonal}.
Although it is not obvious how to disentangle extrinsic and intrinsic influences in complex examples like these, experiments where cells \emph{ex vivo} reproduce aspects of their original population structure \cite{gupta2011state} indicate that there is an intrinsic propensity towards the maintenance of diversity even in absence of external cues.

The dominant explanation is that regulatory feedback loops amplify gene expression noise, eventually giving rise to distinct cells \cite{To2010NoiseCanInduce,elowitz2002science,friedman2006prl,shahrezaei2008pnas,Tan2009Emergent}.
For example, some transcription factors regulate their own expression.
Such a closed circuit may admit more than one stable steady state for appropriate values of the kinetic parameters, which can be occupied by different cells in the population due to the stochastic nature of gene expression, resulting in a bimodal distribution of phenotypes.
These mechanisms are well studied in the literature, with analytical solutions available for their simplest variants \cite{friedman2006prl} and abundant experimental evidence supporting the theoretical results \cite{becskei2000engineering,isaacs2003p,ozbudak2004multistability}.
The main ingredients common to these approaches are: a properly tuned gene regulatory circuit, and gene expression noise.

However, in many contexts partition errors during cell division are more relevant than gene expression noise \cite{huh2011non,huh2011random,bergmiller2017biased,sutterlin2002cell,barrett2017biased,Guantes2015,Chang17}.
Up to now partition noise, which is in average symmetric, has not been studied as a potential source of bimodality in a cell population.
The novelty of this contribution is to propose such a model and justify its relevance in actual biological scenarios.
In particular, we study the division of a cell that carries a certain number of components that influence its growth rate.
Stochastic models of partitioning errors have until now assumed that the growth rate of cells is homogeneous \cite{huh2011non, huh2011random, bergmiller2017biased, sutterlin2002cell, barrett2017biased, Guantes2015, Chang17}.
As we show here, relaxing this assumption is key to obtain a bimodal distribution.


\section*{Model definition}

For simplicity we focus our attention on a single component. Here the component may represent an organelle (e.g., mitochondria) or a macromolecule (e.g., chromosome or plasmid), that will be referred as the component or the particle. We model the particle copy number dynamics across the population as a type-dependent branching process where individual cells replicate at a rate \(\mu_n\), where \(n\) is the number of particles in the cell at birth. To keep things simple, we consider that during their life cycle cells that were born with $k$ particles will duplicate their content resulting $2k$ particles at the time of division. This is a plausible hypothesis for most cell components, which replicate autonomously in coordination with the cell cycle \cite{jajoo2016science}. Notice that this condition could be relaxed and the qualitative picture stays the same . Therefore, a cell born with $k$ copies replicates giving birth to daughter cells with $n$ copies with probability $\Omega_{nk}$. We denote by \(x_n(t)\) the expected value at time $t$ of cells born with particle copy number \(n\). For large populations $x_n(t)$ satisfies the dynamical equation
\begin{equation}
  \dot{x} = W x
  \label{eq:dX}
\end{equation}
where
\begin{equation}
 W = \left(2\Omega - I\right)U
 \label{eq:W}
\end{equation}
$I$ is the identity matrix and $U$ is a diagonal matrix with entries $U_{nn} = \mu_n$. In the long time limit $x(t) \approx c e^{\lambda t}$, where $\lambda$ is the largest eigenvalue of \(W\) and \(c\) its corresponding eigenvector (see Appendix)
\begin{equation}
  \lambda c = W c
  \label{eq:lambda}
\end{equation}
When \(\lambda>0\) the population follows balanced growth and \(\lambda\) represents the average population growth rate \cite{bell1967bio}. 

The forms of \(\mu_n\) and \(\Omega_{nk}\) will depend on specific biological mechanisms controlling growth optimality and partitioning at cell division. We focus our attention on a scenario where molecular mechanisms enforce the maintenance of an optimal growth state with \(n=m\) particles. The enforcement acts at two levels. First, cells will tend to arrest or slow down growth when \(n\) deviates from \(m\). For the sake of illustration and mathematical simplicity we will model the growth rate by the Gaussian function form
\begin{equation}
\mu_n(m,\kappa) = \mu_m \exp\left( - \frac{(n - m)^2}{2\kappa} \right)
\label{eq:mu}
\end{equation}
where \(\mu_m\) is the maximum growth rate and \(\kappa\) quantifies the range of tolerated deviations from the optimal copy number \(m\). Second, cells will tend to enforce even partition at division but, due to stochastic errors, random partition may occur. We introduce a parameter $\epsilon$ between 0 and 1 to quantify the rate of partition errors. Tightly regulated error-free divisions resulting in even distribution of cellular contents correspond to $\epsilon=0$, while absence of control or bias corresponds to $\epsilon=1$, where each particle can be in either daughter cell with equal probability. To interpolate between both situations in a simple manner, we employ the following error prone even partition model,
\begin{equation}
\Omega_{nk}(\epsilon) = \sum_{i=0}^k B_{ik}(\epsilon)B_{n-k+i,2i}(1/2)
\label{eq:even}
\end{equation}
where \(B_{nk}(p)\) denotes the binomial distribution,
\begin{equation}
  B_{nk}(p) = \binom{k}{n} p^n (1-p)^{k-n}
\end{equation}
and \(\epsilon\) is the error probability of random partition per particle pair. Notice that introducing the $\epsilon$ dependency of $\Omega_{nk}$ in this manner, we get an identity matrix if $\epsilon=0$, and a binomial law if $\epsilon=1$, thus smoothly interpolating between the extreme situations just described.
By an appropriate choice of the time unit we can set $\mu_m = 1$ without loss of generality. The model is left then with three parameters, \(\kappa\), \(m\) and \(\epsilon\).


\section*{Results}

\begin{figure*}[!tb]
\centerline{\includegraphics[width=\textwidth]{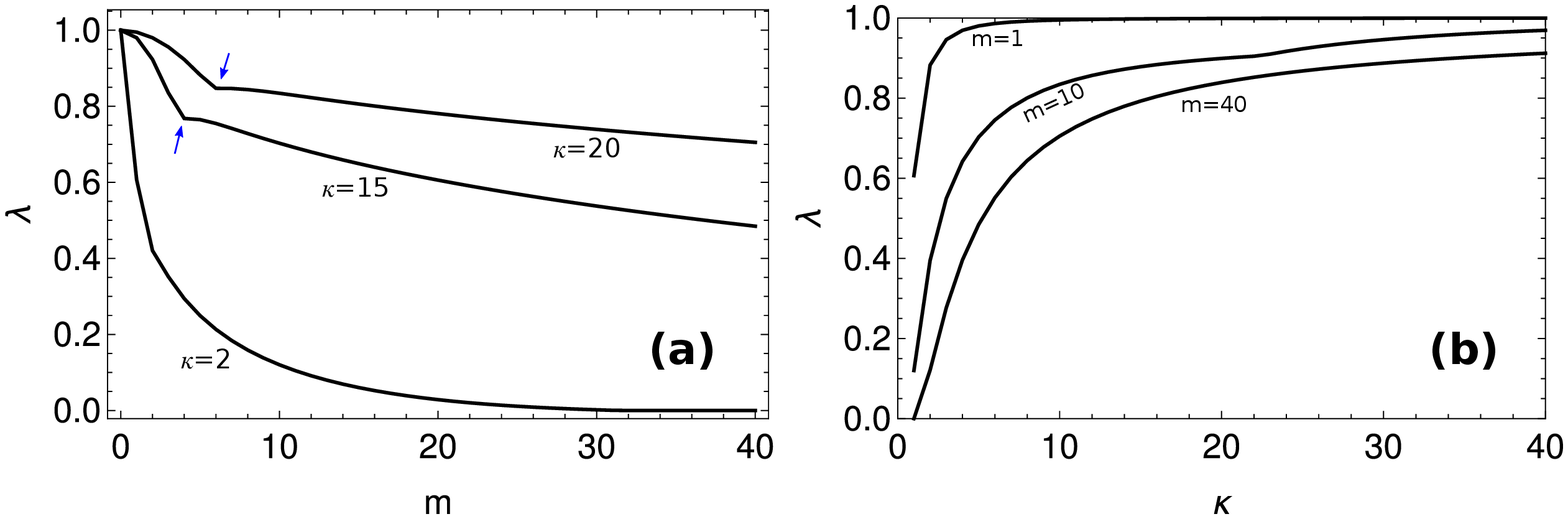}}
\caption{\textbf{Population growth rate.}
(a) Population growth rate as a function of \(m\) for different values of $\kappa$. Here $\epsilon=1$. The arrows indicate inflection points in the dependence of $\lambda$ on $m$. (b) Population growth rate as a function of $\kappa$ for different values of $m$.
\label{fig:lambda}}
\end{figure*}

\begin{figure*}[!t]
\centerline{\includegraphics[width=\textwidth]{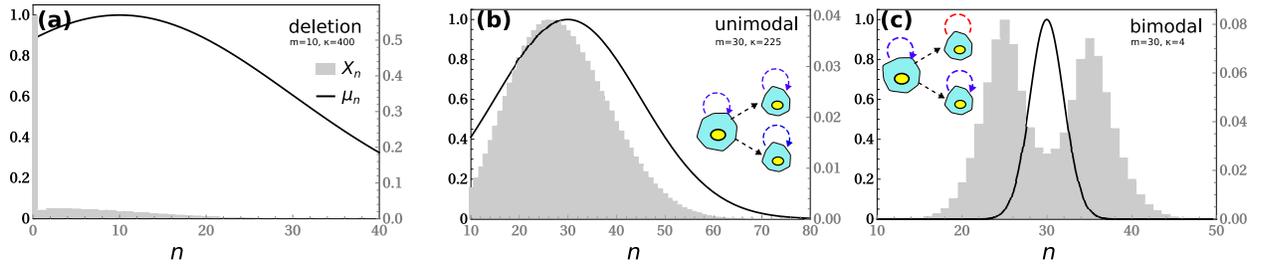}}
\caption{\textbf{Steady state distributions}
Typical steady state distributions for the qualitatively different solutions of the model. The gray bars are the histogram of cells by copy number ($x_n$, right vertical axis), and the dashed curve is the shape of the growth rate as a function of copy number ($\mu_n$, left vertical axis). (a) \emph{Deletion}, where the copy number goes to zero, (b) \emph{unimodal}, where the cell population is distributed in a single bell-like curve around the fitness peak, and (c) \emph{bimodal}, where a significant fraction of newborns fall outside the fitness peak. The inset cartoon diagrams represent the population structures in the unimodal and bimodal regimes.
\label{fig:distributions}}
\end{figure*}

To start our analysis we numerically estimated the population growth rate as a function of \(m\) and \(\kappa\) for the case \(\epsilon=1\). The population growth rate becomes effectively zero when $m$ increases, specially if $m \gg \kappa$ (Fig. \ref{fig:lambda}a). In this limit partition errors drive the majority of cells away from the fitness peak, at the expense of a decreasing pool of fitter cells. In contrast, if $\kappa$ is large then cells are more robust to variations in copy numbers and therefore less sensitive to partition errors (Fig. \ref{fig:lambda}b). This explains the increase in the growth rate with \(\kappa\). It is also evident that \(\lambda\) may exhibit abrupt changes when the parameters vary  (Fig. \ref{fig:lambda}a, blue arrows). This observation suggests that varying the parameter values one may find a solution space characterized by qualitatively different phases.
Indeed, the numerically estimated eigenvector displays different behaviors (Fig. \ref{fig:distributions}). Depending on the choice of the parameters we obtain a population of cells where the particles effectively disappear (Fig. \ref{fig:distributions}a), a homogeneous population with a unimodal distribution of particle copy number (Fig. \ref{fig:distributions}b) or a bimodal distribution of particle copy number (Fig. \ref{fig:distributions}c).
The transition from unimodal to bimodal is continuos as the partition errors increase, with an initial single mode smoothly splitting into two peaks on both sides of the optimal copy number.

To obtain a qualitative insight into the origin of the transition between the different behaviors, we derive approximate analytical solutions.
In the unimodal phase, simulations suggest that the population distribution $c_n$ has a single-peaked bell-like shape that can be approximated by a normal distribution.
This suggests an \emph{ansatz} of the form:
\begin{equation}
  c_n \approx \frac{1}{\sqrt{2\pi\nu}}\exp\left(-\frac{(n-a)^2}{2\nu}\right)
  \label{eq:cn-gaussian}
\end{equation}
with parameters $a,\nu$.
Under this \emph{ansatz} the product $c_n\mu_n$ is also normal.
If we equate the first two moments of the left and right-hand sides of Eq. \eqref{eq:lambda} and take the continuous limit, we obtain a pair of equations that can be solved for $a,\nu$, obtaining (see Appendix for details)
\begin{equation}
  a=m,\quad \nu=(1+\sqrt{1+2r})\kappa/r.
\end{equation}
where
\begin{equation}
r=\frac{2\kappa}{\epsilon m}
\label{eq:r}
\end{equation}
is the ratio between the variances of $\mu_n$ and $\Omega_{mn}$, \emph{i.e.}, the fitness robustness to partition noise ratio. The mean growth rate is approximated by
\begin{equation}
  \lambda_U=\left(\sqrt{1 + \frac{1}{r} + \frac{1}{r}\sqrt{1 + 2r}}\right)^{-1}
\end{equation}
where $\lambda_U$ denotes the value of $\lambda$ in the unimodal phase.
Both $\nu$ and $\lambda_U$ are increasing functions of $r$.

Another possible solution is the deletion phase, where \(c_n=\delta_{n0}\) and \(\lambda_D=\mu_0\). This latter solution will dominate when \(\lambda_D>\lambda_U\) and therefore the condition
\begin{equation}
\mu_0 = \lambda_U(r_{DU})
\label{eq:DU}
\end{equation}
defines the boundary separating the unimodal and deletion phases.

Finally, the numerical simulations in Fig. \ref{fig:distributions}c indicate the existence of bimodal solutions.
One way to understand the emergence of this transition, is to refine the unimodal \emph{ansatz} by further applications of the matrix $W$.
Denote by $c^{(\ell+1)} = W c^{(\ell)}/||W c^{(\ell)}||$ the normalized vector obtained after $\ell+1$ applications of $W$, starting from the unimodal \emph{ansatz} found above, that we now denote by $c^{(0)}$.
As $\ell\rightarrow\infty$ the vector $c^{(\ell)}$ converges to the true eigenvector \cite{demmel1997applied}.
To obtain a tractable analytical expression, we analyze the vector obtained after the first iteration, $c^{(1)}$. Using \eqref{eq:lambda} we can compute the moments of $c^{(1)}$ from those of $c^{(0)}$. In particular, we obtain an expression for the kurtosis of $c^{(1)}$
\begin{equation}
  \begin{aligned}
  K &= \frac{\langle(n-m)^4\rangle_c}{\langle(n-m)^2\rangle_c^2}
  \approx \frac{\langle(n-m)^4\rangle_{c^{(1)}}}{\langle(n-m)^2\rangle_{c^{(1)}}^2} \\
  &= \frac{6 \epsilon}{\left(1 + \sqrt{1+2r} \right) m} - \frac{6
  \epsilon m + 6 \epsilon^2 + 6 \epsilon - 4}{\epsilon m \left( 1 +
  \sqrt{1 + 2 r} \right)^2} + 3
  \end{aligned}
\end{equation}
In the limit $m,\kappa\rightarrow\infty$ with $r=2\kappa/\epsilon m$ fixed, the kurtosis simplifies to
\begin{equation}
  K \sim 3 - \frac{6}{\left( 1 + \sqrt{1 + 2 r} \right)^2}
  \label{eq:Klim}
\end{equation}
which increases with $r$.
The kurtosis measures the \emph{tailedness} of a distribution. In particular, the inequality $K\ge9/5$ holds for all unimodal symmetric distributions \cite{klaassen2000kurtosis}. Though this is only a necessary condition for unimodality (but not sufficient), violation of this inequality can be used as an indication that the unimodality to bimodality transition (BU) has ocurred. From \eqref{eq:Klim} we see that for large $m,\kappa$, this occurs when $r=r_\textrm{BU}\approx0.26$.
An excess of partitioning noise over the robustness of growth is the cause of the bimodal transition in this approximation.

\begin{figure*}[!t]
\centerline{\includegraphics[width=0.5\textwidth]{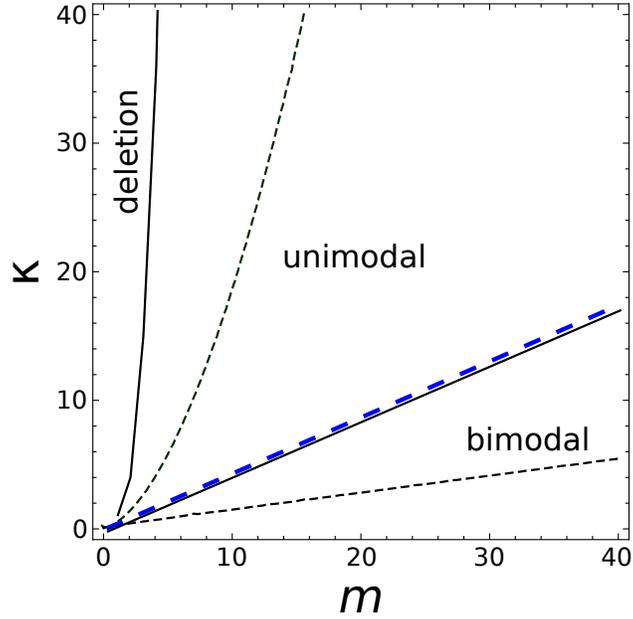}}
\caption{\textbf{Phase diagram} 
Phase plane $\kappa$ vs. $m$ showing regions with different kinds of steady states and the boundary lines separating them, for $\epsilon=1$ (binomial partitioning). Continuous boundaries were obtained from the numerical simulation of the model. Dashed boundaries are the analytical approximation. The dashed blue line is the boundary obtained from equating the kurtosis to 2.15.
\label{fig:phase}}
\end{figure*}

Putting all together, the model can be described by a phase diagram in the $(m,\kappa)$ plane for any particular value of $\epsilon$. In Fig. \ref{fig:phase} we display the result for the bimodal partitioning model, when $\epsilon=1$. Though the analytical approximation described above fails to describe the quantitative location of the boundaries, it captures some of its qualitative features, such as the dependance of the unimodal-bimodal transition on the ratio between $m$ and $\kappa$ and that the regimes $\kappa\gg m$ ($m\gg\kappa$) result in the deletion (bimodal) phases.
 In particular, the bimodal region in the analytical approximation is contained in its simulated counterpart because the inequality $K\le9/5$ is only necessary for unimodality. A corrected threshold $K=2.15$ gives a much better quantitative agreement (see blue dashed line in the figure).
The regions delimited by these boundaries correspond with the following steady states or phases. In the deletion phase phase, the particles effectively disappear from most cells in the population. Since cells cannot synthesize \emph{de novo} particles in this model, the deletion state is irreversible (also called an absorbing state in the language of branching processes). The unimodal phase is characterized by a rather homogeneous population  where the distribution of particle copy number is unimodal. In this case, all cells replicate with appreciable rate and have very similar off-springs, that in turn replicate at similar rates. The third and most relevant phase is characterized by a bimodal distribution of particle copy number. In this case, a minority of cells replicates at an appreciable rate, but because of larger partition errors, the majority of their offspring have too low or too high copy numbers (thus feeding the two peaks with newborn cells) with negligible replication rates.

\begin{figure*}[!t]
\centerline{\includegraphics[width=\textwidth]{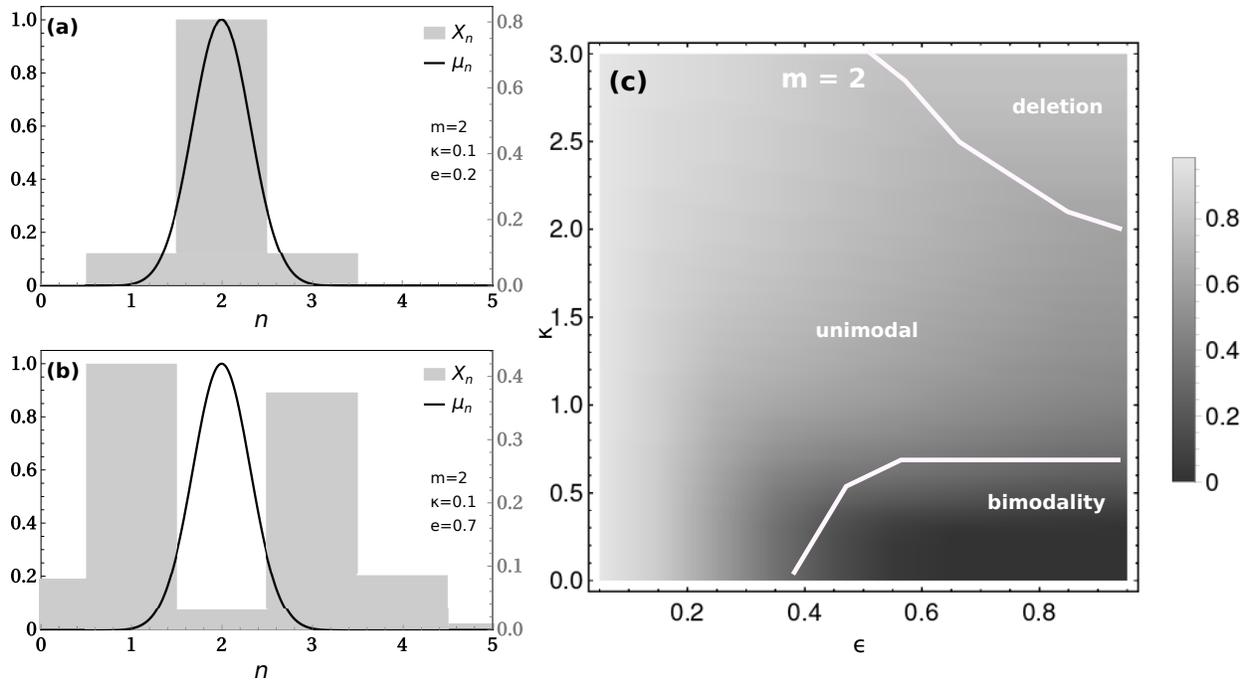}}
\caption{\textbf{Model results for $\epsilon<1,m=2$}, illustrating the unimodal (a) and bimodal regimes (b). (c) Phase diagram $\epsilon$ vs. $\kappa$, for $m=2$. The gradient represents the average growth rate of the population ($\lambda$).
\label{fig:phase_e}}
\end{figure*}

A case of particular interest is that of low $m$ that may represent the evolution of chromosome copy number in a population (ploidy). In the low $m$ limit the analytical approximations are less accurate and we resort on numerical simulations alone. In Fig. \ref{fig:phase_e} we report examples and the phase diagram in the plane \((\epsilon,\kappa)\) for the case $m=2$. The specific examples presented in Fig.  \ref{fig:phase_e}a and b show that for low $m$ there are unimodal and bimodal solutions, respectively. In fact,  the phase diagram retains the three phases: deletion, unimodal and bimodal distributions. It is evident that bimodal solutions are obtained only for partition errors above the threshold $\epsilon_c\approx 0.4$. A second requirement for bimodality is that the growth rate tolerance parameter lies below the threshold $\kappa_c\approx 0.6$, which is equivalent to requirement of small $r$ (\ref{eq:r}) in the phase diagram of Fig. \ref{fig:phase}.

\section*{Discussion}


Mathematical models of partition errors at cell division typically assume that the growth rate of cells is a constant \cite{huh2011non,huh2011random,mukherji2014mechanisms}. This assumption simplifies conveniently mathematical derivations and might be applicable in some particular scenarios. However, as we have shown here, when the growth rate dependency on the copy number is considered, partition errors might be an effective and robust mechanism of cell diversification.

Our  analysis reveals that population bimodality is a feasible state of balanced growth even when all quantities follow unimodal behavior at the single cell level. Within the model considered here, there are two necessary conditions for bimodality to manifest. First, some degree of stochasticity in the partition of particles at division is needed ({\em i.e.}, \(\epsilon>\epsilon_c\)). Second, an optimal copy number and a sharp decrease in the growth rate for sub-optimal copy numbers ({\em i.e.}, small $\kappa$).

At first sight it might seem puzzling that sharp fitness peaks result in bimodality.
One might expect that such a peak would eradicate any deviation from the optimal copy number and thus not allow bimodality.
However, this is why partition errors are essential, because as the fitness peaks get sharper, the probability that a reproductive event results in a newborn inside the peak decreases.
In simple terms, there is a critical point where more individuals are born outside the peak than inside, which results in bimodality in our model.
A similar transition occurs at the error threshold of the quasi-species model \cite{eigen1989molecular}.



There are different biological scenarios where this picture could be relevant. One particularly interesting example is the evolution of ploidy in a population with faulty chromosome segregation discussed above.
Aneuploidy, is a phenomena at the basis of many disabilities and connected with tumorigenesis.
Excluding sex chromosomes, mammalian cells have the diploid state as default, and deviations from this state are rarely tolerated. They have also molecular mechanisms to enforce even chromosome partition at cell division, but mutations in components of the chromosome segregation machinery may lead to an increase of partitioning errors.
However, there is a clear contradiction between the low growth rate of aneuploid cells \cite{Sheltzer12} when compared with normal ones, and their prevalence in the context of cancer.
The theory proposed here provides a hypothesis for how cells with decreased and increased copy number could be more abundant than the ''optimal'' diploid cells.

On the other hand, mitochondria are organelles involved in the production of ATP through oxidative phosphorylation. Unsurprisingly then, experimental evidence shows that the larger the number of mitochondria, the larger the growth rate of cells \cite{Guantes2015}. However, on the extreme case of an excessive mitochondrial content, the decrease of the cytoplasmic space available for other essential components, such as enzymes and ribosomes \cite{vazquez2017, fernandez2017limits, fernandez2018physical} becomes detrimental for the cell. In between there must exist an optimal copy number of mitochondria that maximizes growth in a given environment. However cells typically depart from this optimum. In particular cell differentiation often involves qualitative changes in the quantity of mitochondria content in different stages \cite{katajisto2015science, folmes2012metabolic}. Moreover, recent experiments suggest that the partition of mitochondria at cell division is well approximated by binomial statistics \cite{Guantes2015,jajoo2016science}, but it is not understood how this could support heterogeneous populations. 

Plasmids are small, circular DNA molecules within bacterial cells that replicate autonomously. The abundance of plasmids in a cell is known to affect its growth rate \cite{bentley1990plasmid}, due to excess metabolic burden \cite{paulsson1998jmb} or because the plasmid contains genes that might increase bacterial fitness \cite{maclean2015microbial}. An intermediate plasmid copy number maximizes the growth rate. However, heterogeneity in plasmid content is an important source of evolutionary innovation in bacteria \cite{maclean2015microbial}, but we don't know under which conditions this heterogeneity could be maintained. Partition errors at cell division have been associated with the loss of plasmids in bacterial populations \cite{summers1991kinetics, nordstrom1980partitioning}, in accord with the deletion phase of our model. For example, malfunction of $\mathrm{ParA}_\mathrm{F}$, a protein involved in the regulation of the faithful partition of DNA in bacteria, leads to an enhanced loss rate of plasmids \cite{le2016bacterial}.

Future experimental work is required to validate or invalidate the relevance of the proposed theory in these scenarios.
But our model displays, based on simple and biologically plausible hypothesis, all the phenomenology described by the current experimental results.

Several simplifications were made in the formulation of this model.
We considered an oversimplified view of cell growth by duplication of components without acknowledging regulatory mechanisms coupling component synthesis to cell size and division \cite{taheri2015cell}. 
There is evidence that some components are produced in a cell size dependent manner maintaining constant concentrations \cite{jajoo2016science}, and that cell size homeostasis in a population is achieved by a nearly constant addition of volume after birth (the ``adder'' mechanism) \cite{cadart2018size}.
Investigating the heterogeneity induced by partitioning noise in light of these observations is an interesting topic for a future work.
The ideal Gaussian form of the growth rate \eqref{eq:mu} simplifies our calculations, but it is only an approximation to a generic unimodal bell-like shape expected for a component exhibiting an optimal copy number \cite{paulsson1998jmb, maclean2015microbial, Guantes2015, vazquez2017}.
Partition errors are also subject to increasingly detailed mechanisms depending on the molecule in question, with particular statistical properties \cite{huh2011random} that are not all covered by \eqref{eq:even}.
Inclusion of all these details would have made our model too complicated.
Future work is needed to decide which of these phenomena are more relevant to the mechanism of cell diversification by partitioning errors and growth optimality proposed in this work.

\section*{Acknowledgements}

This project has received funding from the European Union’s Horizon 2020 research and innovation programme MSCA-RISE-2016 under grant agreement No. 734439 INFERNET.

\section*{Author contributions}

JFC, RM and AV developed the theory and wrote the manuscript.

\section*{Competing interests}

The authors declare no competing interests.

\clearpage

\bibliographystyle{naturemag}
\bibliography{partition}

\clearpage

\end{document}